\PassOptionsToPackage{unicode}{hyperref}
\PassOptionsToPackage{hyphens}{url}
\documentclass[
]{article}
\usepackage{lmodern}
\usepackage{amsmath}
\usepackage{ifxetex,ifluatex}
\ifnum 0\ifxetex 1\fi\ifluatex 1\fi=0 % if pdftex
  \usepackage[T1]{fontenc}
  \usepackage[utf8]{inputenc}
  \usepackage{textcomp} % provide euro and other symbols
  \usepackage{amssymb}
\else % if luatex or xetex
  \usepackage{unicode-math}
  \defaultfontfeatures{Scale=MatchLowercase}
  \defaultfontfeatures[\rmfamily]{Ligatures=TeX,Scale=1}
\fi
\IfFileExists{upquote.sty}{\usepackage{upquote}}{}
\IfFileExists{microtype.sty}{% use microtype if available
  \usepackage[]{microtype}
  \UseMicrotypeSet[protrusion]{basicmath} % disable protrusion for tt fonts
}{}
\makeatletter
\@ifundefined{KOMAClassName}{% if non-KOMA class
  \IfFileExists{parskip.sty}{%
    \usepackage{parskip}
  }{% else
    \setlength{\parindent}{0pt}
    \setlength{\parskip}{6pt plus 2pt minus 1pt}}
}{% if KOMA class
  \KOMAoptions{parskip=half}}
\makeatother
\usepackage{xcolor}
\IfFileExists{xurl.sty}{\usepackage{xurl}}{} % add URL line breaks if available
\IfFileExists{bookmark.sty}{\usepackage{bookmark}}{\usepackage{hyperref}}
\hypersetup{
  pdftitle={An Introduction to DoSStoolkit},
  pdfauthor={Rohan Alexander; Samantha-Jo Caetano; Haoluan Chen; Michael Chong; Annie Collins; Shirley Deng; Isaac Ehrlich; Paul Hodgetts; Yena Joo; Marija Pejcinovska; Mariam Walaa; Matthew Wankiewicz},
  hidelinks,
  pdfcreator={LaTeX via pandoc}}
\usepackage[margin=1in]{geometry}
\usepackage{longtable,booktabs}
\usepackage{calc} % for calculating minipage widths
\usepackage{etoolbox}
\makeatletter
\patchcmd\longtable{\par}{\if@noskipsec\mbox{}\fi\par}{}{}
\makeatother
\IfFileExists{footnotehyper.sty}{\usepackage{footnotehyper}}{\usepackage{footnote}}
\makesavenoteenv{longtable}
\usepackage{graphicx}
\makeatletter
\def\maxwidth{\ifdim\Gin@nat@width>\linewidth\linewidth\else\Gin@nat@width\fi}
\def\maxheight{\ifdim\Gin@nat@height>\textheight\textheight\else\Gin@nat@height\fi}
\makeatother
\setkeys{Gin}{width=\maxwidth,height=\maxheight,keepaspectratio}
\makeatletter
\def\fps@figure{htbp}
\makeatother
\providecommand{\tightlist}{%
  \setlength{\itemsep}{0pt}\setlength{\parskip}{0pt}}
\ifluatex
  \usepackage{selnolig}  % disable illegal ligatures
\fi
\newlength{\cslhangindent}
\newlength{\csllabelwidth}
\newenvironment{CSLReferences}[2] % #1 hanging-ident, #2 entry spacing
 {% don't indent paragraphs
  \setlength{\parindent}{0pt}
  \ifodd #1 \everypar{\setlength{\hangindent}{\cslhangindent}}\ignorespaces\fi
  \ifnum #2 > 0
  \setlength{\parskip}{#2\baselineskip}
  \fi
 }%
 {}
\usepackage{calc}

\title{An Introduction to \texttt{DoSStoolkit}\thanks{We gratefully acknowledge a grant from the University of Toronto Faculty of Arts \& Science Pedagogical Innovation and Experimentation Fund. We thank Bethany White, Liza Bolton, Monica Alexander, Radu Craiu, and Sabrina Sixta for their support. Correspondence: \href{mailto:rohan.alexander@utoronto.ca}{\nolinkurl{rohan.alexander@utoronto.ca}}}}
\usepackage{etoolbox}
\makeatletter
\providecommand{\subtitle}[1]{% add subtitle to \maketitle
  \apptocmd{\@title}{\par {\large #1 \par}}{}{}
}
\makeatother
\subtitle{Interactive, student-developed, self-paced, modules for learning R}
\author{Rohan Alexander \and Samantha-Jo Caetano \and Haoluan Chen \and Michael Chong \and Annie Collins \and Shirley Deng \and Isaac Ehrlich \and Paul Hodgetts \and Yena Joo \and Marija Pejcinovska \and Mariam Walaa \and Matthew Wankiewicz}
\date{19 May 2021}

\begin{document}
\maketitle
\begin{abstract}
We describe a series of interactive, student-developed, self-paced, modules for learning R. We detail the components of this resource, and the pedagogical underpinning. We discuss the development of this resource, and avenues for future work. Our resource is available as an R package: \texttt{DoSStoolkit}
\end{abstract}

\hypertarget{introduction}{%
\section{Introduction}\label{introduction}}

A team consisting of faculty, graduate, and undergraduate students from the Department of Statistical Sciences (DoSS) at the University of Toronto has developed a series of interactive, self-paced, modules for learning the statistical programming language R (R Core Team 2020). These modules are based on the \texttt{learnr} R package (Schloerke et al. 2021) and available for public use at: \url{http://dosstoolkit.com/}.

The use-case of the \texttt{DoSStoolkit} is to support the teaching and learning of R programming (as well as some complementary aspects, for instance GitHub and R Markdown) that occurs both in the classroom and online. R is not an explicit requirement in any particular course, however it is used in almost all of the courses offered in the Department of Statistical Sciences (both graduate and undergraduate). These courses serve roughly four thousand undergraduate students, and 100 graduate students within the department; as well as hundreds of others (undergraduates, graduates, instructors, etc.) outside of the department. Some students, particularly those without any experience with programming languages, struggle as much, if not more, with R compared to statistics. However, there is not time to explicitly teach all aspects of R that a course will draw on. And in any case, it is generally the case that programming languages, for our purposes, are best taught within the context of being used to achieve some project, rather than in and of themselves. Hence, the need for this \texttt{DoSStoolkit}.

A user can get started with the \texttt{DoSStoolkit} using RStudio Cloud, and instructions are included in the package website. The first module is largely motivational. Additionally, the first module goes through installing R and RStudio locally, which enables the user to transition away from RStudio Cloud before exceeding the free trial period. That said, given the variety of student computational set-ups, \texttt{DoSStoolkit} testing has focused on RStudio Cloud and the University of Toronto JupyterHub. There are ten modules in all, which broadly cover: dealing with errors; organizing code and folders; the essential \texttt{dplyr} verbs (Wickham et al. 2021); base functions; dealing with data; \texttt{ggplot} (Wickham 2016); \texttt{R\ Markdown} (Xie, Dervieux, and Riederer 2020); GitHub; iteration, packages, and \texttt{Shiny} (Chang et al. 2021).

There are three distinguishing and innovative features of the \texttt{DoSStoolkit}. Firstly, almost all materials were developed by undergraduate students. This provides the materials with a `voice' that students who are new to R may be better able to relate to. It also ensures that the materials focus on what those who have recently learnt R think is useful. Secondly, the materials integrate videos, examples, common errors, as well as formative assessment. However, the modules of the \texttt{DoSStoolkit} are completely self-paced and able to be used by students on an as-needed basis, compared with a traditional textbook or course notes that may be accessed in more of a linear fashion. Finally, the materials are focused on the realities of using R as an applied statistician or applied statistics student. For instance, there are entire lessons about using Stack Overflow, and Googling, as well as a focus on the community aspect of R.

We have found that the \texttt{DoSStoolkit} supports student learning during our initial use-cases and are looking to continue to develop and iterate on it. We are also looking to further integrate it into classes. Our materials are available under an MIT license, and we expect that many other high schools and universities will look to build out their own versions of these materials.

The remainder of this paper is structured as follows: Section 2 details the content that is covered; Section 3 discusses the pedagogical underpinnings for those interested in using the \texttt{DoSStoolkit} in their own teaching; Section 4 is a broad discussion section about the development, assessment, and other points, as well as detailing the planned next steps.

\hypertarget{description-and-content}{%
\section{Description and content}\label{description-and-content}}

The \texttt{DoSStoolkit} is divided into ten modules that each focus on a particular aspect of R programming. In this section we will describe the main aspects of each module. To begin, a new R learner needs to install various packages as well as \texttt{DoSStoolkit}. We expect new R learners are using either RStudio Cloud, or the University of Toronto JupyterHub, and so expect at least R 4.0.3.

\hypertarget{hello-world}{%
\subsection{Hello world!}\label{hello-world}}

The first module focuses on downloading R and RStudio and understanding the key aspects, such as the console, script editor, file and plot viewer, and environment panes. The user is then encouraged to get familiar with R and RStudio by stepping through a few exercises. These exercises are at a level that a user may expect to be after a few weeks of experience. They are designed to be aspirational and encouraging. These exercises consist of tasks where users need to simply copy-paste code. This makes the exercises easily achievable from the get-go and provides the user with a sense of success and encouragement early on. These exercises do not provide a deep understanding of what is going on, but instead are intended to help ease the user into programming. Finally, this module covers a variety of aspects of the R community that a new user may be interested to be part of. For instance, the R Weekly newsletter, R Ladies groups, and R meetups.

The full list of lessons in this module is:

\begin{itemize}
\tightlist
\item
  Why I love R, by Liza Bolton.
\item
  Setting up RStudio, by Annie Collins.
\item
  Getting to know what is what - console, terminal, etc, by Annie Collins.
\item
  A fun hello world exercise, by Annie Collins.
\item
  Another fun hello world exercise, by Shirley Deng.
\item
  R Weekly newsletter, R Ladies, R meetups, by Annie Collins.
\end{itemize}

\hypertarget{operating-in-an-error-prone-world}{%
\subsection{Operating in an error prone world}\label{operating-in-an-error-prone-world}}

A common issue for new R users is that they are able to run code that was delivered in class, but run into trouble when they try to change any aspect of this. The second module first normalises the need to get help and then works through strategies for solving problems. These include advice for using Google and Stack Overflow, problem solving, how to ask for help and creating reproducible examples, and finally how to make the most of the error messages that are present.

One important aspect in this module is that the user is given code that deliberately generates errors and is then asked, in a controlled way, to fix it. This is part of reminding new users that errors are a normal, common part of using R and errors are in no way a reflection of the user being `bad at programming,' but instead are a common obstacle that we all face and need to work through. If a writer makes a typo, then this doesn't necessarily make them a bad writer because there are tools such as spellcheck, and processes such as re-reading, that will hopefully enable them to fix the typo and continue writing. This module is designed to provide new R learners with analogous strategies and tools so that they can address their own issues and carry on.

The full list of lessons in this module is:

\begin{itemize}
\tightlist
\item
  Why I love R, by Monica Alexander.
\item
  Getting help is normal, by Michael Chong.
\item
  Using Google and Stack Overflow, by Michael Chong.
\item
  Stack overflow, by Annie Collins.
\item
  How to problem solve when your code doesn't work, by Michael Chong.
\item
  Making reproducible examples, by Marija Pejcinovska.
\item
  How to make the most of R's cryptic error messages, by Shirley Deng.
\end{itemize}

\hypertarget{holding-the-chaos-at-bay}{%
\subsection{Holding the chaos at bay}\label{holding-the-chaos-at-bay}}

The third module has to do with organization of both the code itself that is written and the project structure. We emphasize how to set up R projects, and why they are important, and provide some recommended folder set ups. Examples of written comments are provided and then packages are introduced. Finally, this module covers common ways of reading in data.

The full list of lessons in this module is:

\begin{itemize}
\tightlist
\item
  Why I love R, by Samantha-Jo Caetano.
\item
  R projects and \texttt{setwd()}, by Isaac Ehrlich.
\item
  Folder set-up, by Isaac Ehrlich.
\item
  Writing comments, by Isaac Ehrlich.
\item
  \texttt{install.packages()}, by Haoluan Chen.
\item
  \texttt{install\_github()}, by Haoluan Chen.
\item
  \texttt{library()}, by Mariam Walaa.
\item
  \texttt{update.packages()}, by Mariam Walaa.
\item
  \texttt{read\_csv()}, by Marija Pejcinovska.
\item
  \texttt{read\_table()}, \texttt{read\_dta()}, and other data types, by Isaac Ehrlich.
\end{itemize}

\hypertarget{hand-me-my-plyrs}{%
\subsection{Hand me my plyrs}\label{hand-me-my-plyrs}}

The fourth module focuses on \texttt{dplyr} and covers a lot of core material that are used extensively in DoSS courses, mostly within \texttt{tidyverse} (Wickham et al. 2019). For instance, the pipe, and the main \texttt{dplyr} verbs: \texttt{select()}, \texttt{filter()}, \texttt{group\_by()} and \texttt{ungroup()}, \texttt{summarise()}, \texttt{arrange()}, \texttt{mutate()}, \texttt{rename()}, \texttt{count()}, and \texttt{slice()}. This module also contains lessons on functions that are commonly used alongside these core verbs including \texttt{pivot\_wider()} and \texttt{pivot\_longer()}. Finally, it contains lessons about data types - character, numeric, dates, logical - and the various ways in which datasets are commonly considered in R, including vectors, matrices, dataframes, and tibbles.

The full list of lessons in this module is:

\begin{itemize}
\tightlist
\item
  Why I love R, by Sabrina Sixta.
\item
  What is the tidyverse?, by Yena Joo.
\item
  The pipe, by Mariam Walaa.
\item
  \texttt{select()}, by Yena Joo.
\item
  \texttt{filter()}, by Shirley Deng.
\item
  \texttt{group\_by()} and \texttt{ungroup()}, by Matthew Wankiewicz.
\item
  \texttt{summarise()}, by Mariam Walaa.
\item
  \texttt{arrange()}, by Isaac Ehrlich.
\item
  \texttt{mutate()}, by Haoluan Chen.
\item
  \texttt{pivot\_wider()} and \texttt{pivot\_longer()}, by Annie Collins.
\item
  \texttt{rename()}, by Mariam Walaa.
\item
  \texttt{count()} and \texttt{uncount()}, by Annie Collins.
\item
  \texttt{slice()}, by Annie Collins.
\item
  \texttt{c()}, \texttt{matrix()}, \texttt{data.frame()}, and \texttt{tibble()}, by Matthew Wankiewicz.
\item
  \texttt{length()}, \texttt{nrow()}, and \texttt{ncol()}, by Isaac Ehrlich.
\end{itemize}

\hypertarget{totally-addicted-to-base}{%
\subsection{Totally addicted to base}\label{totally-addicted-to-base}}

The fifth module focuses on base functions. In general, most DoSS classes are based in the \texttt{tidyverse} environment, however given the statistical content, there is also a considerable amount of base that is used. These include summary statistics, but also more broad programming concepts such as functions, for and while loops, if, and also some base graphing options.

The full list of lessons in this module is:

\begin{itemize}
\tightlist
\item
  Why I love R, by Rohan Alexander.
\item
  \texttt{mean()}, \texttt{median()}, \texttt{sd()}, \texttt{lm()}, and \texttt{summary()}, by Mariam Walaa.
\item
  \texttt{function()}, by Haoluan Chen.
\item
  \texttt{for()}, and \texttt{while()}, by Yena Joo.
\item
  \texttt{if()}, \texttt{if\_else()}, and \texttt{case\_when()}, by Haoluan Chen.
\item
  \texttt{c()}, \texttt{seq()}, \texttt{seq\_along()}, and \texttt{rep()}, by Matthew Wankiewicz.
\item
  \texttt{hist()}, \texttt{plot()}, and \texttt{boxplot()}, by Yena Joo.
\end{itemize}

\hypertarget{he-was-a-d8er-boi}{%
\subsection{He was a d8er boi}\label{he-was-a-d8er-boi}}

Data are central to many of the classes at DoSS and so we devote an entire module to this. To a large extent this involves aspects of data cleaning, preparation, and cleaning that are typically not aspects that a student would be explicitly taught. This includes dealing with strings, joining datasets, identifying missing data, simulating data, across, working with factors, dates, and regular expressions. This is a large and varied module, but the expectation is that a student could complete specific lessons without needing to complete the entire module, filling in gaps as needed.

The full list of lessons in this module is:

\begin{itemize}
\tightlist
\item
  \texttt{head()}, \texttt{tail()}, \texttt{glimpse()}, and \texttt{summary()}, written by Haoluan Chen.
\item
  \texttt{paste()}, \texttt{paste0()}, \texttt{glue::glue()} and \texttt{stringr}, written by Marija Pejcinovska
\item
  \texttt{names()}, \texttt{rbind()}, and \texttt{cbind()}, written by Isaac Ehrlich.
\item
  \texttt{left\_join()}, \texttt{anti\_join()}, \texttt{full\_join()}, etc, written by Haoluan Chen.
\item
  Looking for missing data, written by Mariam Walaa.
\item
  \texttt{set.seed()}, \texttt{runif()}, \texttt{rnorm()}, and \texttt{sample()}, written by Haoluan Chen.
\item
  Simulating datasets for regression, written by Mariam Walaa.
\item
  Advanced mutating and summarising, written by Mariam Walaa.
\item
  Tidying up datasets, written by Mariam Walaa.
\item
  \texttt{pull()}, \texttt{pluck()}, and \texttt{unnest()}, by Isaac Ehrlich.
\item
  \texttt{forcats} and factors, written by Matthew Wankiewicz.
\item
  More on strings, written by Annie Collins.
\item
  Regular expressions, written by Shirley Deng.
\item
  Working with dates, written by Mariam Walaa.
\item
  \texttt{janitor}, written by Mariam Walaa.
\item
  \texttt{tidyr}, written by Mariam Walaa.
\end{itemize}

\hypertarget{to-ggplot-or-not-to-ggplot}{%
\subsection{To ggplot or not to ggplot}\label{to-ggplot-or-not-to-ggplot}}

We focus an entire module on the use of \texttt{ggplot} due to its importance in an applied statistics workflow. We begin by explaining broadly what \texttt{ggplot} is and its role. We then move to discussing different options depending on the situation: one categorical variable; one continuous variable; two variables; geographic variables.

The full list of lessons in this module is:

\begin{itemize}
\tightlist
\item
  \texttt{ggplot2::ggplot()}, by Shirley Deng.
\item
  Bar charts, by Matthew Wankiewicz.
\item
  Histograms, by Haoluan Chen.
\item
  Scatter plots, by Haoluan Chen.
\item
  Various useful options, by Yena Joo.
\item
  Saving graphs, by Yena Joo.
\end{itemize}

\hypertarget{r-marky-markdown-and-the-funky-docs}{%
\subsection{R Marky Markdown and the Funky Docs}\label{r-marky-markdown-and-the-funky-docs}}

Increasingly DoSS expects students to write R Markdown documents and submit either the .Rmd itself or a document generated from it. This module details all aspects of R Markdown in a practical sense. For instance, detailing various top matter opinions such as title and date; options around tables and graphs; and referencing.

The full list of lessons in this module is:

\begin{itemize}
\tightlist
\item
  Introduction to R Markdown, written by Shirley Deng.
\item
  Top Matter: Title, Date, Author, Abstract, written by Yena Joo.
\item
  Tables: \texttt{kable}, \texttt{kableextra}, and \texttt{gt}, written by Yena Joo.
\item
  Multiple plots with \texttt{patchwork}, written by Michael Chong
\item
  References and Bibtex, written by Yena Joo.
\item
  PDF outputs, written by Yena Joo.
\item
  \texttt{here::here()} and filepaths, written by Matthew Wankiewicz.
\end{itemize}

\hypertarget{git-outta-here}{%
\subsection{Git outta here}\label{git-outta-here}}

Git, GitHub, and version control is increasingly integrated into the work of applied statisticians, and its use in DoSS has been particularly helpful for group work during the pandemic. Although Git, GitHub and version control have been well integrated into applied statistics, many undergraduate students have little experience with it. In this module, we introduce version control and GitHub and then a workflow that might be appropriate for someone working by themselves or perhaps with another person i.e.~pull, status, add, commit, and push. We then add another layer, in terms of branches and conflicts. We focus on using this in the context of RStudio, rather than command line. While that removes some of the power of git, we hope that it provides some comfort and familiarity to new R learners.

The full list of lessons in this module is:

\begin{itemize}
\tightlist
\item
  What is version control and GitHub?, written by Mariam Walaa.
\item
  Git: pull, status, add, commit, push, written by Mariam Walaa.
\item
  Branches in GitHub, written by Matthew Wankiewicz.
\item
  Dealing with Conflicts, written by Matthew Wankiewicz.
\item
  Putting (G)it All together in RStudio, written by Matthew Wankiewicz.
\end{itemize}

\hypertarget{indistinguishable-from-magic}{%
\subsection{Indistinguishable from magic}\label{indistinguishable-from-magic}}

One of the exciting aspects of R is how it is possible to quickly achieve relatively sophisticated outcomes. This module covers iteration, writing R packages, creating websites using \texttt{Blogdown} (Xie, Dervieux, and Hill 2021), and building apps using \texttt{Shiny} (Chang et al. 2021). We also include a lesson on improving R coding skills. Although the modules are designed to be self-contained and able to be used in a piecemeal fashion, if a student were to have gone through the \texttt{DoSStoolkit} in a linear fashion, by the time they are at this module they have a thorough foundation of R skills. If they had at the same time, been completing projects and using R for classes, they could quite reasonably describe themselves as an intermediate R user. They would have sufficient experience and skills to be able to work as a teaching assistant for introductory R courses.

The full list of lessons in this module is:

\begin{itemize}
\tightlist
\item
  Coding style, written by Marija Pejcinovska.
\item
  Static maps with \texttt{ggmap}, by Annie Collins.
\item
  Writing R Packages, written by Matthew Wankiewicz.
\item
  Getting started with \texttt{Blogdown}, written by Annie Collins.
\item
  Getting started with \texttt{Shiny}, written by Matthew Wankiewicz.
\end{itemize}

\hypertarget{pedagogical-underpinning}{%
\section{Pedagogical underpinning}\label{pedagogical-underpinning}}

From a pedagogical perspective the \texttt{DoSStoolkit} poses many advantages and builds on considerable pedagogical research.

\hypertarget{students-teaching-students}{%
\subsection{Students-teaching-students}\label{students-teaching-students}}

The first is that the majority of the materials were created by a team of undergraduate students who would have only been exposed to R in the past year. Having these students take on the teacher role, as students-teaching-students, has been shown to be pedagogically ideal (Beasley 1997), (Wagner and Gansemer-Topf 2005), (Stigmar 2016). Compared with instructors, the student-teachers creating these modules are likely more aware of the common questions and misunderstandings among new R learners and are better able to develop materials that will pre-emptively fill knowledge gaps. Additionally, the role of students as the teacher is likely better received by the new learner (Stigmar 2016). To a new learner, it may be daunting to learn from, and have to reach out to, an instructor, making the end goal of learning seem further away or less attainable. Whereas, if a recent R learner is the person instructing, then the end goal may seem more manageable and the idea of asking for help when needed perhaps more comfortable.

Wilson (2019) says `the biggest motivators for adult learners are their sense of agency (i.e., the degree to which they feel that they're in control of their lives), the utility or usefulness of what they're learning, and whether their peers are learning the same things.' We hope that having these modules written by fellow students who are also learning about these R tools is useful in this regard. We also are deliberately trying to create an accepted set of R knowledge that was deliberately chosen and created, as opposed to other tools a student might independently find but that may not be as helpful.

\hypertarget{equitable-experience}{%
\subsection{Equitable experience}\label{equitable-experience}}

An additional pedagogical benefit of the \texttt{DoSStoolkit} is the parallel benefit that it allows for an equitable experience to most users. The nature of the modules being completable asynchronously allows for flexibility in the new R learners' schedule. Thus, the \texttt{DoSStoolkit} is accessible to new R learners of different groups that may normally be disadvantaged due to scheduling issues (learners in different time-zones, learners who are parents, learners who work, etc.) (Gillis and Krull 2020). With that being said, we do recognize the benefits of incorporating synchronous components to promote student engagement and improve the overall learning experience (Littenberg-Tobias and Reich 2020). The intent is that synchronous components can be implemented by DoSS instructors when using the \texttt{DoSStoolkit} in their respective statistics and data science courses accordingly, based on the structure of their course assessments (e.g., having weekly quizzes based on \texttt{DoSStoolkit} modules).

The toolkit, although developed to be aligned with DoSS courses, is free and accessible to the public. Additionally, the \texttt{DoSStoolkit} modules consist of both text and video content. The text content is readable by most audio transcribing software (Seale 2013). The video components are accompanied by closed captioning text (Seale 2013).

\hypertarget{end-user-benefits}{%
\subsection{End user benefits}\label{end-user-benefits}}

Although, the intended purpose of the \texttt{DoSStoolkit} is to support the supplementary R coding materials in senior level applied DoSS courses, the \texttt{DoSStoolkit} will be freely available and accessible to anyone. This was done for two main reasons. The first, is that we recognize that statistics and data science are ubiquitous to most science research, thus there is a need for R learning in other departments outside of statistics. The second reason is that there is a need for R learning for learners who are not current (or prospective) undergraduate level students. This rationale, paired with the formative assessments also allow the end users to use the \texttt{DoSStoolkit} to fill knowledge gaps as needed. Additionally, the \texttt{DoSStoolkit} paired with the summative assessments could be used by a prospective faculty member who wishes to learn R to teach it to their students.

\hypertarget{enhanced-learning}{%
\subsection{Enhanced learning}\label{enhanced-learning}}

One reason for developing the \texttt{DoSStoolkit} is because new R learners have described how difficult they found it to find resources for learning R that include everything they needed to know. They described how they often ended up having multiple tabs open from a variety of different websites, trying to figure out how to do something. When they are successful, the result is often a collection of code that despite working is likely to be brittle. And they are often not successful, leading to frustration and feelings of not being capable. No single course teaches R, but it is largely assumed. The result has been that students learn it as they go, which is understandably difficult and frustrating.

\hypertarget{discussion}{%
\section{Discussion}\label{discussion}}

We see the \texttt{DoSStoolkit} as establishing a platform on which we can build. In particular, the online nature of the \texttt{DoSStoolkit} enables experimentation in terms of developing best-practice; the self-paced learning aspect enables us to offer courses to a wider variety of students than might otherwise take courses in the department; and finally, the open source and permissive license enables our work to be adapted by others to their circumstances.

\hypertarget{development-and-updating}{%
\subsection{Development and updating}\label{development-and-updating}}

The initial development of the \texttt{DoSStoolkit} was conducted over an eight-month period. As the majority of the content was written by undergraduate students, much of the work occurred in just two one-month periods following the Fall 2020 and Winter 2021 terms. Due to the pandemic, all work was conducted fully remotely.

An initial call for undergraduate research assistants was established and this yielded hundreds of applications. Applicants were selected based on samples of R code that they provided through GitHub, recommendations, but most importantly, signals that the student was interested in communicating and sharing their knowledge. Students were deliberately selected from a variety of experience levels but given the way that the university is set-up the majority of students were in third and fourth year.

We also recruited two DoSS graduate students. These students wrote content that required more experience, and also acted as mentors and a conduit between the two faculty PIs and the undergraduates. Finally, they also helped develop the broader framework for the \texttt{DoSStoolkit}, taking advantage of their experience both learning and teaching at DoSS, providing experience that the PIs did not have.

The undergraduates were given some direction and initial training on \texttt{learnr}, but in general were given, and encouraged to take advantage of, considerable latitude when developing their materials. The faculty PIs and graduate students established a list of topics and the undergraduates selected ones that they were particularly interested in, and the remainder were assigned.

The \texttt{DoSStoolkit} received funding of \$15,000 from the Faculty of Arts and Sciences fund designed to encourage pedagogical innovation. These funds were spent on wages for the graduate and undergraduate students. The average wage was almost \$30 per hour. Similar \texttt{learnr} modules could be established with a relatively small seed. We were able to accomplish much of what we set out to achieve with the first \$5,000 and the remaining funds allowed us to build out much more. One small aspect to be aware of is that there are considerable pressures on undergraduate students these days and in general this was not the type of project that many of them had time for during term, and so structuring the development timeline is important. We found that one- or two-week `sprints' before term started, and after exam periods, were particularly useful. The highly selected nature of the undergraduate students involved meant that a large amount could be achieved in a relatively short period.

The modules have been created as an R Package that is hosted on GitHub. This allows the use of GitHub issues for requests and minor changes as students go through the materials. Although this has not yet been accomplished, we plan to submit this package to CRAN to ease installation. Finally, the \texttt{learnr} package that underpins \texttt{DoSStoolkit} is still under considerable development. We found breaking changes and considerable differences in the ability to run code between different machines. Ambitious projects, such as \texttt{DoSStoolkit}, will require a considerable amount of maintenance and updating, requiring some on-going funds.

\hypertarget{assessment}{%
\subsection{Assessment}\label{assessment}}

Although the focus of the \texttt{DoSStoolkit} is that it complements courses taught in DoSS and those courses would have their respective assessments, we have integrated a great deal of formative assessments throughout almost all of the lessons. In general, this is done using multi-choice questions, but fill-in-the-blank and other assessments are also used. A student is prompted to actively work to satisfy these formative assessment items. For instance, they may be expected to write out some code or adapt some other code. Initial feedback is that we need to build out considerably more formative assessment, as new R learners have found this particularly useful.

One important aspect of this formative assessment is to get students used to working out what is wrong. Wilson (2019) says `It is almost oxymoronic to say that learners spend a lot of their time trying to figure out what they've done wrong and fixing it: after all, if they knew and they had, they would already have moved on to the next subject.' More-experienced R users know what to do to work through an error, however new R learners may just feel frustrated. The hope is that these types of formative assessment items help students learn how to identify and deal with errors

At the moment there is no summative assessment in the \texttt{DoSStoolkit}. However, we intend to build this out in future iterations. We expect summative assessment would be useful for two main reasons. Firstly, it would encourage students to complete multiple lessons and modules, hopefully building out their knowledge in a broader way than might otherwise happen. Secondly, a particular level of accomplishment could be used as a signalling device from students to teachers that they have a certain level of knowledge. The inverse situation has already begun to happen in DoSS, where professors specify a certain level of knowledge as a prerequisite for a course and are using the \texttt{DoSStoolkit} to signal these requirements.

Summative assessment is being developed. Our current approach is to associate levels with a set of different colours: white, yellow, orange, green, purple, blue, brown, and black. Each colour draws from lessons across various modules, so it is not necessary to have completed the entire module in order to undertake the assessment. White is designed to be achievable with only an hour or two of effort and essentially just involves installing R and RStudio locally and running through some vignettes. The higher levels add on various skills and expectations, with black being associated with having completed the teaching modules and is something that DoSS would expect of teaching assistants. However, summative assessment remains an area for future development.

\hypertarget{other-points}{%
\subsection{Other points}\label{other-points}}

\hypertarget{experimentation}{%
\subsubsection{Experimentation}\label{experimentation}}

The \texttt{DoSStoolkit} establishes a platform for experiments. While courses have a chance to develop over time as an instructor refines them, they are typically not subject to experiments to see what works. Partly this is appropriate as it would be difficult to assess in an experimental way. However, this is also partly due to the difficulty of randomisation. The \texttt{DoSStoolkit} enables instructors to randomize different explanations and assessments, to better understand what might be effective in terms of student learning.

\hypertarget{diversity-in-statistical-sciences}{%
\subsubsection{Diversity in statistical sciences}\label{diversity-in-statistical-sciences}}

It is well-known that statistical sciences suffer from a lack of diversity. While it still occurs, it is most pronounced in PhD and faculty ranks, rather than at an undergraduate level. Statistics courses can be intimidating for students to start. By allowing students to learn the essentials of R at their own pace, it is hoped that the \texttt{DoSStoolkit} will enable a wider variety of students to succeed in statistics courses and continue onto graduate studies and faculty appointments.

\hypertarget{other-applications}{%
\subsubsection{Other applications}\label{other-applications}}

Similar methods and approaches could be established for a wide variety of aspects of statistical sciences education. For instance, learning about distributions and randomization, as well as the essentials of proofs are two areas that come to mind as potentially fruitful avenues for development. Additionally, it would make sense to develop a companion `textbook' in a similar way to the \texttt{learnr} modules of Çetinkaya-Rundel (2021), and Field (2020) supplement a text.

\hypertarget{concluding-remarks}{%
\subsubsection{Concluding remarks}\label{concluding-remarks}}

We hope that the \texttt{DoSStoolkit} makes R more appealing, accessible, and approachable. We also hope that others build on, and adapt, this foundation for their own purposes. The nature of applied statistics has changed quickly with the rise of data science, and we expect there to be considerable further work along these lines in the future.

\newpage

\hypertarget{references}{%
\section*{References}\label{references}}
\addcontentsline{toc}{section}{References}

\hypertarget{refs}{}
\begin{CSLReferences}{1}{0}
\leavevmode\hypertarget{ref-beasley1997students}{}%
Beasley, Colin. 1997. {``Students as Teachers: The Benefits of Peer Tutoring.''} \emph{Learning Through Teaching}, 21--30.

\leavevmode\hypertarget{ref-citeshiny}{}%
Chang, Winston, Joe Cheng, JJ Allaire, Carson Sievert, Barret Schloerke, Yihui Xie, Jeff Allen, Jonathan McPherson, Alan Dipert, and Barbara Borges. 2021. \emph{Shiny: Web Application Framework for r}. \url{https://CRAN.R-project.org/package=shiny}.

\leavevmode\hypertarget{ref-citedsbox}{}%
Çetinkaya-Rundel, Mine. 2021. \emph{Dsbox: Data Science Course in a Box}.

\leavevmode\hypertarget{ref-citeadventr}{}%
Field, Andy. 2020. \emph{Adventr: Interactive r Tutorials to Accompany Field (2016), "an Adventure in Statistics"}. \url{https://CRAN.R-project.org/package=adventr}.

\leavevmode\hypertarget{ref-gillis2020covid19}{}%
Gillis, Alanna, and Laura M Krull. 2020. {``COVID-19 Remote Learning Transition in Spring 2020: Class Structures, Student Perceptions, and Inequality in College Courses.''} \emph{Teaching Sociology} 48 (4): 283--99.

\leavevmode\hypertarget{ref-littenberg2020evaluating}{}%
Littenberg-Tobias, Joshua, and Justin Reich. 2020. {``Evaluating Access, Quality, and Equity in Online Learning: A Case Study of a MOOC-Based Blended Professional Degree Program.''} \emph{The Internet and Higher Education} 47: 100759.

\leavevmode\hypertarget{ref-citeR}{}%
R Core Team. 2020. \emph{{R: A Language and Environment for Statistical Computing}}. Vienna, Austria: R Foundation for Statistical Computing. \url{https://www.R-project.org/}.

\leavevmode\hypertarget{ref-citelearnr}{}%
Schloerke, Barret, JJ Allaire, Barbara Borges, and Garrick Aden-Buie. 2021. \emph{{learnr: Interactive Tutorials for R}}.

\leavevmode\hypertarget{ref-seale2013learning}{}%
Seale, Jane. 2013. \emph{E-Learning and Disability in Higher Education: Accessibility Research and Practice}. Routledge.

\leavevmode\hypertarget{ref-stigmar2016peer}{}%
Stigmar, Martin. 2016. {``Peer-to-Peer Teaching in Higher Education: A Critical Literature Review.''} \emph{Mentoring \& Tutoring: Partnership in Learning} 24 (2): 124--36.

\leavevmode\hypertarget{ref-wagner2005learning}{}%
Wagner, Mimi, and Ann Gansemer-Topf. 2005. {``Learning by Teaching Others: A Qualitative Study Exploring the Benefits of Peer Teaching.''} \emph{Landscape Journal} 24 (2): 198--208.

\leavevmode\hypertarget{ref-citeggplot}{}%
Wickham, Hadley. 2016. \emph{Ggplot2: Elegant Graphics for Data Analysis}. Springer-Verlag New York. \url{https://ggplot2.tidyverse.org}.

\leavevmode\hypertarget{ref-citetidyverse}{}%
Wickham, Hadley, Mara Averick, Jennifer Bryan, Winston Chang, Lucy D'Agostino McGowan, Romain François, Garrett Grolemund, et al. 2019. {``Welcome to the {tidyverse}.''} \emph{Journal of Open Source Software} 4 (43): 1686. \url{https://doi.org/10.21105/joss.01686}.

\leavevmode\hypertarget{ref-citedplyr}{}%
Wickham, Hadley, Romain François, Lionel Henry, and Kirill Müller. 2021. \emph{Dplyr: A Grammar of Data Manipulation}. \url{https://CRAN.R-project.org/package=dplyr}.

\leavevmode\hypertarget{ref-wilson2019ten}{}%
Wilson, Greg. 2019. {``Ten Quick Tips for Creating an Effective Lesson.''} \emph{PLoS Computational Biology} 15 (4): e1006915.

\leavevmode\hypertarget{ref-citeblogdown}{}%
Xie, Yihui, Christophe Dervieux, and Alison Presmanes Hill. 2021. \emph{Blogdown: Create Blogs and Websites with r Markdown}. \url{https://github.com/rstudio/blogdown}.

\leavevmode\hypertarget{ref-citermarkdown}{}%
Xie, Yihui, Christophe Dervieux, and Emily Riederer. 2020. \emph{R Markdown Cookbook}. Boca Raton, Florida: Chapman; Hall/CRC. \url{https://bookdown.org/yihui/rmarkdown-cookbook}.

\end{CSLReferences}

\end{document}